\documentclass[prl,comment,twocolumn,showpacs]{revtex4}

\usepackage{amsmath}
\usepackage{amsfonts}
\usepackage{amssymb}
\usepackage [ansinew] {inputenc}
\usepackage{natbib}

\usepackage{theorem}
\theorembodyfont{\upshape}

\newcommand{\J} {
\ensuremath{\mathcal{J}}
}

\begin{document}
\author{Vladimir Garc\'{\i}a-Morales}
\email{vladimir.garcia@uv.es}
\author{Javier Cervera}
\title{Comment on ``Is Tsallis Thermodynamics Nonextensive?'' by E. Vives and A. Planes [cond-mat/0106428]}
\affiliation{Department of Thermodynamics, University of Valencia, 
E-46100 Burjassot, Spain}
\begin{abstract}
We comment on letter ``Is Tsallis Thermodynamics Nonextensive?'' by E. Vives and A. Planes
[Phys. Rev. Lett. \textbf{88}, 020601 (2002) cond-mat/0106428]. 
It is pointed out that the Euler and Gibbs-Duhem equations derived in the
letter can serve to justify an appropriate form for the Lagrange parameters
controlling thermal equilibrium, without need of any change of variables. 
This leads to a framework for Tsallis Thermodynamics
which is free from recent criticisms raised by Nauenberg [Phys. Rev. E \textbf{67}, 036114 (2003)
cond-mat/0210561] and Gross [Physica (Amsterdam) \textbf{305}, 99 (2002) cond-mat/0106496]. 
This is accomplished through a direct connection with Hill's Nanothermodynamics.
\end{abstract}
\pacs{05.70.Ln, 05.20.Gg, 05.40.-a}
\maketitle

In a recent letter \cite{Vives} Vives and Planes (VP) discuss 
Tsallis Thermodynamics (TT) deriving 
a generalization of the Gibbs-Duhem (GD) equation 
which reduces to the traditional one when the entropic parameter $q$ 
tends to unity. The authors suggest a change of variables 
that allows to recover standard Thermodynamics, proposing expressions
for the Lagrange parameters (LPs) which are supposed to be those controlling
mutual equilibrium between thermodynamic systems.
Although the approach leading to Eq. (9) in \cite{Vives} is correct and insightful, 
the change of variables and subsequent analysis is unnecessary 
and leads to controversial analytical expressions for the LPs
that have been the kernel of the main criticisms
raised against TT \cite{Nauenb}. We point out here that Eqs. (4), (8) and (9) in \cite{Vives} 
can serve as a basis to justify that the LPs 
$y_{\alpha}$ in \cite{Vives} are the \emph{physically meaningful ones}
contrarily to what is suggested by the authors and other practitioners in the field 
(see Refs. $[12]$ and $[13]$ in \cite{Vives}). 
This leads us to establish equilibrium properties for TT overcoming
previous difficulties \cite{Nauenb}. The $y_{\alpha}$ can be viewed 
as the LPs controlling (nano)thermodynamic equilibrium
and, contrarily to what is claimed by the authors, these are then intensive variables. 
To proceed further let us introduce Hill's formalism of Nanothermodynamics (NT)\cite{Hill0,Hill}. 
Hill's NT is a rigorous extension of standard thermodynamics to systems
that experience equilibrium fluctuations of arbitrary strenght 
(``small systems'') in which all quantities involved have a clear physical meaning. 
We provide next a connection between TT and NT. It is interesting to note that the (entropic) 
Euler and GD equations in NT (see Eqs. (1-72) and (1-75) in Ref. \cite{Hill}) are, respectively
(rewritten here in VP notation)
\begin{eqnarray}
\sum_{\alpha}y_{\alpha,H}\left<X_{\alpha}\right>_{H}=S-\J \label{HEuler}\\
-\sum_{\alpha}\left<X_{\alpha}\right>_{H}dy_{\alpha,H}=d\J \label{HGD}
\end{eqnarray}
Subindex $H$ means ``Hill's variables'' which are the physical (averaged) extensive 
($\left<X_{\alpha}\right>$) and intensive ($y_{\alpha}$) ones. 
$S$ is the physical entropy for 
\emph{one} system and $\J$ is the subdivision (entropic) potential. 
$\J$ is an intensive variable and the number of systems $\lambda$ 
is its conjugate extensive one (in our case, as in \cite{Vives}, we are dealing with only
one system and $\lambda=1$ is implicitly considered in these equations). 
By differentiating Eq.(\ref{HEuler}) and using Eq.(\ref{HGD})
it can be seen that the entropy for one system $S$ satisfies the differential equation 
$dS=\sum_{\alpha}y_{\alpha,H}d\left<X_{\alpha}\right>_{H}$
which is formally identical to Eq.(4) in \cite{Vives}. 
It is clear that the structure of VP equations
and those of Hill is the same and the former can be obtained from the latter 
if $S \equiv \mathcal{S}^{*}$, $y_{\alpha,H} \equiv y_{\alpha}$, 
$\left<X_{\alpha}\right>_{H}\equiv\left<X_{\alpha}\right>_{q}$ and 
the following correspondence is made
\begin{equation}
\J \equiv \mathcal{S}^{*}-\left[1+(1-q)\mathcal{S}^{*}\right]\frac{\ln \left[1+(1-q)\mathcal{S}^{*}\right] }{1-q}
\label{def}
\end{equation}
Here Tsallis entropy is not only the physical one: its 
nonextensivity property (see Eq. (2) in \cite{Vives}) is also the basis for the 
subdivision entropic potential $\J$ which is found to be necessary to explain 
the thermal behavior of small systems (at least). TT describes, thus,
the most general thermal equilibrium, the nanothermodynamic equilibrium \cite{Chamber}, 
in which the new potential $\J$ plays a decisive role. $\J$ 
vanishes for a macroscopic (extensive) system, for which one has also $q=1$ in Eq.(\ref{def}),
and is a measure of the degree of fragmentation of a system in smaller (nonextensive) 
subsystems. Through the correspondence established above and from Hill's NT, it is now known
that the $y_{\alpha}$ must be equal for different
systems put in contact at equilibrium. \emph{The $y_{\alpha}$ are then the physically meaningful}
LPs. The main additional feature is
that the potential $\J$ must also be equal for systems at equilibrium. 
This implies in TT that, for two different systems $A$ and $B$
\begin{eqnarray}
&&\mathcal{S}_{A}^{*}-\frac{1}{1-q_{A}}\left[1+(1-q_{A})\mathcal{S}_{A}^{*}\right]
\ln \left[1+(1-q_{A})\mathcal{S}_{A}^{*}\right] \nonumber \\
&&=\mathcal{S}_{B}^{*}-\frac{1}{1-q_{B}}\left[1+(1-q_{B})\mathcal{S}_{B}^{*}\right]
\ln \left[1+(1-q_{B})\mathcal{S}_{B}^{*}\right]\nonumber
\end{eqnarray}
This is to be considered an additional equilibrium condition that nonextensive systems must meet.
When $q_{A}$ and $q_{B}$ tend to unity both sides of this equation vanish
and the standard thermodynamic equilibrium (controlled only by the LPs
$y_{\alpha}$) is regained. This condition, besides some VP results and the connection with NT, 
allows to establish equilibrium TT, which is 
now free from recent criticisms \cite{Nauenb} arising from the use of inappropriate LPs.


\begin{thebibliography}{}

\bibitem{Vives}
E. Vives and A. Planes, Phys. Rev. Lett. \textbf{88}, 020601 (2002) cond-mat/0106428.
\bibitem{Nauenb}
M. Nauenberg, Phys. Rev. E \textbf{67}, 036114 (2003) cond-mat/0210561;
D. H. E. Gross, Physica (Amsterdam) \textbf{305}, 99 (2002) cond-mat/0106496.
\bibitem{Hill0}
T. L. Hill, J. Chem. Phys. \textbf{36}, 3182 (1962).
\bibitem{Hill}
T. L. Hill, \emph{Thermodynamics of Small Systems} (Dover, New York, 1994).
\bibitem{Chamber}
R. V. Chamberlin, Phys. Lett. A \textbf{315}, (3-4), 313 (2003).
\end{thebibliography}
\end{document}